\documentclass[preprint,amsmath,amssymb,aps,superscriptaddress]{revtex4-2}
\usepackage{amsfonts}
\usepackage{stmaryrd}
\usepackage{bbm}
\usepackage{mathrsfs}
\usepackage{tipa}
\usepackage{amssymb}
\usepackage{txfonts}
\usepackage{graphicx}
\usepackage{dcolumn}
\usepackage{epstopdf}
\usepackage[colorlinks,linkcolor=blue,urlcolor=blue,citecolor=blue]{hyperref}
\usepackage{multirow}
\usepackage{subfigure}
\usepackage{url}
\usepackage{upgreek}
\usepackage[utf8]{inputenc}
\usepackage[english]{babel}
\usepackage{lineno}
\usepackage{xcolor}
\usepackage{ulem}

\begin{document}
\title{Collapse of density wave and emergence of superconductivity in pressurized-La$_4$Ni$_3$O$_{10}$ evidenced by ultrafast spectroscopy}

\author{Shuxiang Xu}
\email{shuxiangxu@pku.edu.cn}
\affiliation{International Center for Quantum Materials, School of Physics, Peking University, Beijing 100871, China}

\author{Hao Wang}
\affiliation{International Center for Quantum Materials, School of Physics, Peking University, Beijing 100871, China}

\author{Mengwu Huo}
\affiliation{Guangdong Provincial Key Laboratory of Magnetoelectric Physics	and Devices, School of Physics, Sun Yat-Sen University, Guangzhou, Guangdong 510275, China}

\author{Deyuan Hu}
\affiliation{Guangdong Provincial Key Laboratory of Magnetoelectric Physics	and Devices, School of Physics, Sun Yat-Sen University, Guangzhou, Guangdong 510275, China}

\author{Qiong Wu}
\affiliation{International Center for Quantum Materials, School of Physics, Peking University, Beijing 100871, China}

\author{Li Yue}
\affiliation{Beijing Academy of Quantum Information Sciences, Beijing 100913, China}

\author{Dong Wu}
\affiliation{Beijing Academy of Quantum Information Sciences, Beijing 100913, China}

\author{Meng Wang}
\affiliation{Guangdong Provincial Key Laboratory of Magnetoelectric Physics	and Devices, School of Physics, Sun Yat-Sen University, Guangzhou, Guangdong 510275, China}

\author{Tao Dong}
\email{taodong@pku.edu.cn}
\affiliation{International Center for Quantum Materials, School of Physics, Peking University, Beijing 100871, China}

\author{Nanlin Wang}
\email{nlwang@pku.edu.cn}
\affiliation{International Center for Quantum Materials, School of Physics, Peking University, Beijing 100871, China}
\affiliation{Beijing Academy of Quantum Information Sciences, Beijing 100913, China}
\affiliation{Collaborative Innovation Center of Quantum Matter, Beijing 100871, China}

\date{\today}

\begin{abstract}

Recent discoveries of superconductivity in Ruddlesden-Popper nickelates realize a rare category of superconductors. However, the use of high-pressure diamond anvil cells limits spectroscopic characterization of the density waves and superconducting gaps. Here, we systematically studied the pressure evolution of La$_{4}$Ni$_{3}$O$_{10}$ using ultrafast optical pump-probe spectroscopy. We found that the transition temperature and energy gap of density waves are suppressed with increasing pressure and disappear suddenly near 17 GPa where structural transition appears. In addition, the observation of a single density wave gap indicates that the spin density wave and charge density wave remain coupled as pressure increases, rather than decoupling. After the density wave collapse, a distinct low-temperature regime emerges, characterized by a small gap consistent with potential superconducting pairing. The separated phase region of superconductivity and density waves suggests that superconductivity in pressurized-La$_4$Ni$_3$O$_{10}$ competes strongly with density waves, offering new insights into the interplay between these two phenomena. 

\end{abstract}

\maketitle 
\clearpage

\section{Introduction}
Ruddlesden-Popper nickelates, as high-$T_{c}$ superconductors, have attracted widespread attention since the discovery of their superconducting properties \cite{Sun2023,Zhang2024high,Wang2024,Dong2024}. Like cuprates and iron-based superconductors, superconductivity in nickelates also emerges from a magnetic ground state \cite{Armitage2010,Dai2015,Keimer2015,Chen2024}. In nickelates, spin density wave and charge density wave coexist at low temperature \cite{Liu2022,Zhang2020}, both of which can be suppressed by applying pressure. A pressure-induced phase transition from the density wave state to superconductivity occurs at low temperatures \cite{Sun2023,Zhang2024high,Wang2024,Zhu_2024,Li__2024,zhang2024}. For instance, La$_{4}$Ni$_{3}$O$_{10}$, a prototypical Ruddlesden-Popper nickelate, exhibits concurrent spin density wave (SDW) and charge density wave (CDW) order at $T_{DW}$ = 136 K \cite{Zhang2020}. Under pressures, transport measurements indicate a rapid suppression of $T_{DW}$, then superconductivity (SC) appears after spin and charge density waves (DW) disappear\cite{Li__2024,zhang2024}, or coexists with the DW at low pressures \cite{Zhu_2024}. Notably, recent study on tetragonal La$_{4}$Ni$_{3}$O$_{10}$ suggests that the DW order rather than tetragonal lattice is a prerequisite for realizing SC \cite{shi2025}. These findings raise critical questions regarding the correlation among DW order, lattice distortion and SC. Investigations of the bilayer La$_{3}$Ni$_{2}$O$_{7}$ have revealed a right-angled triangular superconducting phase diagram, indicating that the superconducting transition in La$_{3}$Ni$_{2}$O$_{7}$ is likely first order\cite{Sun2023,li2024pressure}. How this behavior compares with the trilayer compound La$_{4}$Ni$_{3}$O$_{10}$ remains an open question. To elucidate these issues, further spectroscopic studies on pressurized La$_{4}$Ni$_{3}$O$_{10}$ are essential. A comprehensive understanding of the pressure evolution of DW order and the intricate interplay among SDW, CDW, lattice distortion and SC in La$_{4}$Ni$_{3}$O$_{10}$ is crucial for advancing our understanding of nickelate high-$T_{c}$ superconductivity.

Currently, studies on the pressure evolution of DW and SC in La$_{4}$Ni$_{3}$O$_{10}$ are primarily limited on high-pressure electrical transport, while spectroscopic measurement under high pressure is challenging and relatively rare \cite{Zhu_2024,Li__2024,zhang2024}. Techniques such as angle-resolved photoemission spectroscopy and scanning tunneling spectroscopy, which are commonly employed to detect the energy gap of the electronic orders, are infeasible under pressure due to the constraints imposed by sample encapsulation within a diamond anvil cell (DAC). Optical pump–probe spectroscopy, known for its highly sensitive to even subtle charge gaps in the density of states (DOS), has been successfully utilized to investigate various competing orders and extract the corresponding gap values in high-$T_{c}$ superconductors \cite{PhysRevLett.82.4918,PhysRevLett.104.027003,Dong2022} and correlated materials \cite{Giannetti2016,PhysRevB.99.165144}. Recently, ultrafast pump–probe spectroscopy has been employed to probe the electron dynamics of DW in La$_{4}$Ni$_{3}$O$_{10}$ at ambient pressure \cite{LI2025180,xu2024origin} by measuring the amplitude and decay time of the photoexcited quasiparticles (QPs). However, the high-pressure spectroscopic information, particularly the DW and SC gap value under pressure, has yet to be determined. 

In this work, we successfully developed an in-situ low-temperature, high-pressure pump-probe system by combining the pump-probe spectroscopy with low temperature and high pressure technology as shown in the Supplementary Figs.1-3. With high-quality ultrafast dynamic data of La$_{4}$Ni$_{3}$O$_{10}$ under pressure, we found that the $T_{DW}$ first decreases slowly with pressure and then collapses suddenly near 15 GPa where a structural transition occurs \cite{Zhu_2024,Li2024Stru}. Next, a new charge gap, likely related to SC, was observed at low temperature. The energy gaps of DW and SC under various pressures were estimated using the Rothwarf-Taylor (RT) model \cite{PhysRevLett.19.27}. Our results suggested the emergence of SC in La$_{4}$Ni$_{3}$O$_{10}$ is accompanied by the sudden disappearance of the DW, indicating a strongly competitive interplay between superconductivity and density waves.  

\section{Results and discussion}

\begin{figure}
	\includegraphics[width=0.9\columnwidth]{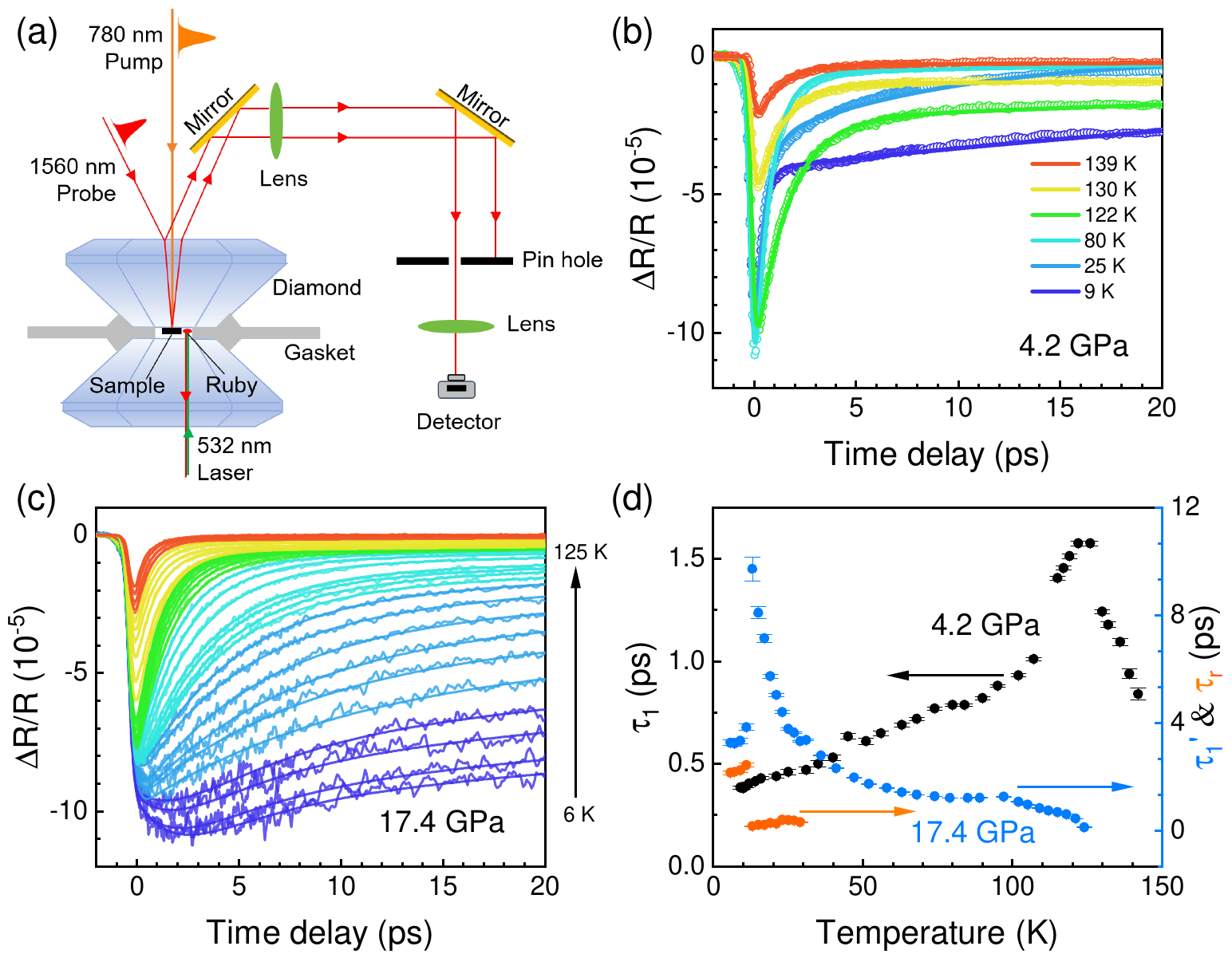}
	\caption{\textbf{The schematic of the in-situ low temperature, high pressure optical pump-probe spectroscopy and the temperature-dependent transient reflectivity $\Delta R/R$ at two representative pressures.}  (a) A schematic illustration of measuring the variation of reflectivity with centre wavelength of 1560 nm after excitation with a central wavelength of 780 nm inside diamond anvil cells. The diamond anvils are of type IIa, with a transmittance exceeding 70 \%. Two lenses are used to distinguish the sample signal from the signal of the diamond's upper surface. (b) The $\Delta R/R$ at typical temperatures and 4.2 GPa. The hollow circle is the raw data, while the solid lines are fitting curves obtained using Eq. (1). (c) The $\Delta R/R$ measured at 17.4 GPa from 6 K to 125 K, with a temperature interval of 2 K in the range of 6 K to 30 K and 5 K in the range of 30 K to 125 K. The smooth solid lines are fitting curves using Eq. (2). Both (b) and (c) were measured at a pump fluence of $F$ = 0.17 $\mu$J/cm$^{2}$. (d) Temperature dependence of the decay time of the reflectivity transients extracted from fits to the double exponential decay function. The black circle refers to $\tau_{1}$ of 4.2 GPa fitted with Eq. (1), while the blue (red) circle is $\tau_{1}^{\prime}$ ($\tau_{r}$) of 17.4 GPa fitted with Eq. (2). Error bars refer to the standard error in the exponential fitting.
	} 
	
	\label{Figure1}
\end{figure}

Fig. \ref{Figure1}(a) shows the schematic of high pressure optical pump-probe measurements in which a 120 fs pump pulse with 1.58 eV is used to excite the pressurized La$_{4}$Ni$_{3}$O$_{10}$ sample and another time-delayed probe pulse at 0.79 eV is used to record the relative differential reflectivity $\Delta R/R$ as a function of delay time after excitation. Figs. \ref{Figure1}(b) and (c) show the transient reflectivity $\Delta R/R$ of two representative pressures of 4.2 GPa and 17.4 GPa, respectively, obtained at a pump fluence of $F$ = 0.17 $\mu$J/cm$^{2}$. At 4.2 GPa, the density wave still survives at low temperatures. The corresponding $\Delta R/R$ at 9 K reaches negative maximum quickly after the photoexcitation at the zero delay time followed by picosecond time scale decay processes. With increasing the temperatures, the decay becomes slower and reaches its slowest at 122 K. Then, the decay reversely becomes faster, and the amplitude decreases rapidly with increasing temperature. To decouple the different decay processes, the transient reflectivity $\Delta R/R$ curves were fitted using a double-exponential function convolved with a Gaussian laser pulse:  
\begin{equation}
	\frac{\Delta R}{R}(t) = \frac{1}{\sqrt{2\pi}\omega}\exp\left(\frac{-t^{2}}{2\omega^{2}}\right)\otimes\left\{A_{1}\exp\left(\frac{-(t-t_{0})}{\tau_{1}}\right) + A_{2}\exp\left(\frac{-(t-t_{0})}{\tau_{2}}\right)\right\}
\end{equation}
where $A_{1}$ and $A_{2}$ are the amplitude, $\tau_{1}$ and $\tau_{2}$ are the relaxation time of the fast and slow decay processes which are often assigned as the electron-phonon and lattice related thermalization process \cite{Giannetti2016}, respectively,  $\omega$ is the temporal duration of the incident pulse. The fast decay component $\tau_{1}$, shown as the black dots in Fig. \ref{Figure1}(d), is strongly temperature-dependent and exhibits a distinct divergent feature near 122 K corresponding to the DW gap opening. Such a divergent feature can be understood using phenomenological RT model, which initially was used to explain the strong phonon bottleneck effect in conventional superconductors \cite{PhysRevLett.19.27}, but now extended its application to other charge gap opening orders such as CDW/SDW, unconventional superconductors \cite{Giannetti2016} and heavy fermion materials \cite{HF_PP}. Following excitation, if there is a small gap ($\Delta$) in the DOS, the recombination of the QPs across the gap will create high frequency ($\omega > 2\Delta$) phonons (HFP). Since these phonons can subsequently excite additional QPs, the ultimate recovery of the photoexcited QPs back to the equilibrium states is governed by the decay of the high frequency phonon population. As the gap closes near transition temperature, more thermally excited low energy phonons become available for QPs production, impeding the recombination of photoexcited QPs and reducing their ratio to thermally excited QPs, which then cause the observed divergence in $ \tau_1$ and sharp dropping in $A_{1}$, respectively. This is the so called phonon bottleneck effect. In the weak perturbation limit, as an sufficiently low fluence used in our measurement, the RT model can be solved analytically and the expressions usually are used to extract the gap value by fitting temperature-dependent $ \tau$ and $A$, when one assumes the gap function below transition temperature is Bardeen-Cooper-Schrieffer (BCS) like. We will discuss the DW gap value ($\Delta_{DW}$) estimation latter. 

With increasing the pressure to 17.4 GPa, the low temperature relaxation process is rather distinct. The divergent feature arising from the phonon bottleneck effect of the DW gap in the relaxation process is indistinguishable as shown in Fig. \ref{Figure1}(c). Instead, a slow rise process becomes evident at low temperatures, which is very similar to that of the superconducting state in MgB$_{2}$ \cite{PhysRevLett.91.267002} and electron-doped cuprate La$_{2-x}$Ce$_{x}$CuO$_{4}$ \cite{PhysRevB.95.115125}. This slow-rise component then makes the fitting by Equation (1) insufficient as shown in Supplementary Fig. 4, an additional term resembling superconductivity-induced rise is necessary to fit the low temperature $\Delta R/R$ transient well:
\begin{equation}
	\frac{\Delta R}{R}(t) = \frac{1}{\sqrt{2\pi}\omega}\exp\left(\frac{-t^{2}}{2\omega^{2}}\right)\otimes M
\end{equation} 
\begin{equation}
	M=A_{1}^{\prime}\exp\left(\frac{-(t-t_{0})}{\tau_{1}^{\prime}}\right)+A_{2}^{\prime}\left(1-\exp\left(\frac{-(t-t_{0})}{\tau_{r}}\right)\right)\exp\left(\frac{-(t-t_{0})}{\tau_{2}^{\prime}}\right)
\end{equation}
The second term on the right side of equation has previously been used to resolve the superconducting signal in optimal doping Nd$_{2-x}$Ce$_{x}$CuO$_{4+\delta}$ \cite{PhysRevLett.110.217002}. In La$_{4}$Ni$_{3}$O$_{10}$, the $\Delta R/R$ doesn't change its sign below the superconducting transition. There are two main reasons why the SC signal in La$_{4}$Ni$_{3}$O$_{10}$ remains negative, unlike that in optimally doped Nd$_{2-x}$Ce$_{x}$CuO$_{4+\delta}$. On the one hand, both Nd$_{2-x}$Ce$_{x}$CuO$_{4+\delta}$ and La$_{2-x}$Ce$_{x}$CuO$_{4}$  contain competing orders (superconductivity and antiferromagnetic fluctuations). The optimally doped La$_{2-x}$Ce$_{x}$CuO$_{4}$ (x=0.11) also displays a noticeable sign change near $T_{c}$\cite{PhysRevB.95.115125}. However, the underdoped La$_{2-x}$Ce$_{x}$CuO$_{4}$ (x=0.08)\cite{PhysRevB.95.115125}, which still has the competing orders, shows that the SC component remains negative, similar to La$_{4}$Ni$_{3}$O$_{10}$. It seems that the slight variation of doping level significantly impacts the behavior of $\Delta R/R$. Therefore, the different SC behavior between La$_{4}$Ni$_{3}$O$_{10}$ and optimally doped Nd$_{2-x}$Ce$_{x}$CuO$_{4+\delta}$ likely originates from the difference in band structures and the corresponding temperature dependence of optical constants. It should be mentioned that the magnetic field could separate the SC component from the total $\Delta R/R$ response at selective temperatures where the 6 T magnetic field is sufficient to suppress the superconductivity completely. The separated SC component indeed is negative in underdoped La$_{2-x}$Ce$_{x}$CuO$_{4}$ (x=0.08) but positive in optimally doped La$_{2-x}$Ce$_{x}$CuO$_{4}$ (x=0.11). In our current setup, the cryostat is too large to be compatible with the magnetic field, so magnetic field-dependent $\Delta R/R$ cannot be obtained. On the other hand, in optimally doped Nd$_{2-x}$Ce$_{x}$CuO$_{4+\delta}$, the SC coexists and competes with antiferromagnetic fluctuations below $T_{c}$. In La$_{4}$Ni$_{3}$O$_{10}$, the DW is suppressed completely above 17.4 GPa. No long-range DW competes with the SC phase in La$_{4}$Ni$_{3}$O$_{10}$, which also may lead to the sign of $\Delta R/R$ remaining intact across superconductivity transition. Thus, careful fitting is necessary to extract the superconductivity signal. The smooth solid lines in Fig. \ref{Figure1}(c) matched the experimental data very well. The blue and red dots in Fig. \ref{Figure1}(d) plot the extracted $\tau_{1}^{\prime}$ and $\tau_{r}$, which respectively show a divergence in $\tau_{1}^{\prime}$ and onset behavior in $\tau_{r}$ at 17 K. The temperature-dependent $\tau_{1}^{\prime}$ also can be understood using RT model. Here the slow rise component $\tau_{r}$ implies that after the excitation, photoexcited QPs initially generate a high density of HFP, which in turn excite the additional QPs across the gap until the quasiequilibrium between QPs and HFP populations is reached. After the quasi-stationary state is established, the decay of the HFP will govern the following relaxation process which result in the divergent feature in $\tau_{1}^{\prime}$. The appearance of $\tau_{r}$ and the phonon bottleneck effect at 17 K lead to an inescapable conclusion that there is a small gap in La$_{4}$Ni$_{3}$O$_{10}$ below 17 K under 17.4 GPa pressure. Next, we consider the physical origin of this small gap above 17.4 GPa. On the one hand, although the optical pump-probe spectroscopy is extremely sensitive to the charge gap opening in the DOS, it is unable to definitively identify which symmetry breaking order, e.g., SDW, CDW or SC, opens the observed gap. Beside the SC, a small SDW/CDW gap also can lead to the similar slow rise feature in the $\Delta R/R$ transients, which has been reported in the pump-probe measurement on URu$_2$Si$_2$ \cite{PhysRevB.84.161101} below its hidden order phase transition temperature. Up to now, no other electronic instabilities except SC has been reported in pressured La$_{4}$Ni$_{3}$O$_{10}$ above 17 GPa by transport measurement. Furthermore, as shown in the Supplementary Fig. 6, the low temperature slow rise component appears below 23 K with pressure up to 25.6 GPa, which is consistent with SC transition temperature determined by high-pressure transport experiment \cite{Zhu_2024,Li__2024,zhang2024}. On the other hand, we measured the relaxation behavior of a pure gasket under pressure and excluded the possibility of gasket signal. The relaxation process of the gasket becomes faster as the temperature decreases, which differs significantly from the signal of the sample, as shown in the Supplementary Fig. 9. In addition, we used the principle of pinhole imaging to separate the signal from the top surface of the diamond. Therefore, the signal we measured here originates entirely from the sample. These results allow us to conclude that it is most likely a small superconducting gap leading to the observed low temperatures slow rise component at high pressures. We will extract the SC gap value ($\Delta_{SC}$) latter.

\begin{figure}
	\includegraphics[width=0.95\columnwidth]{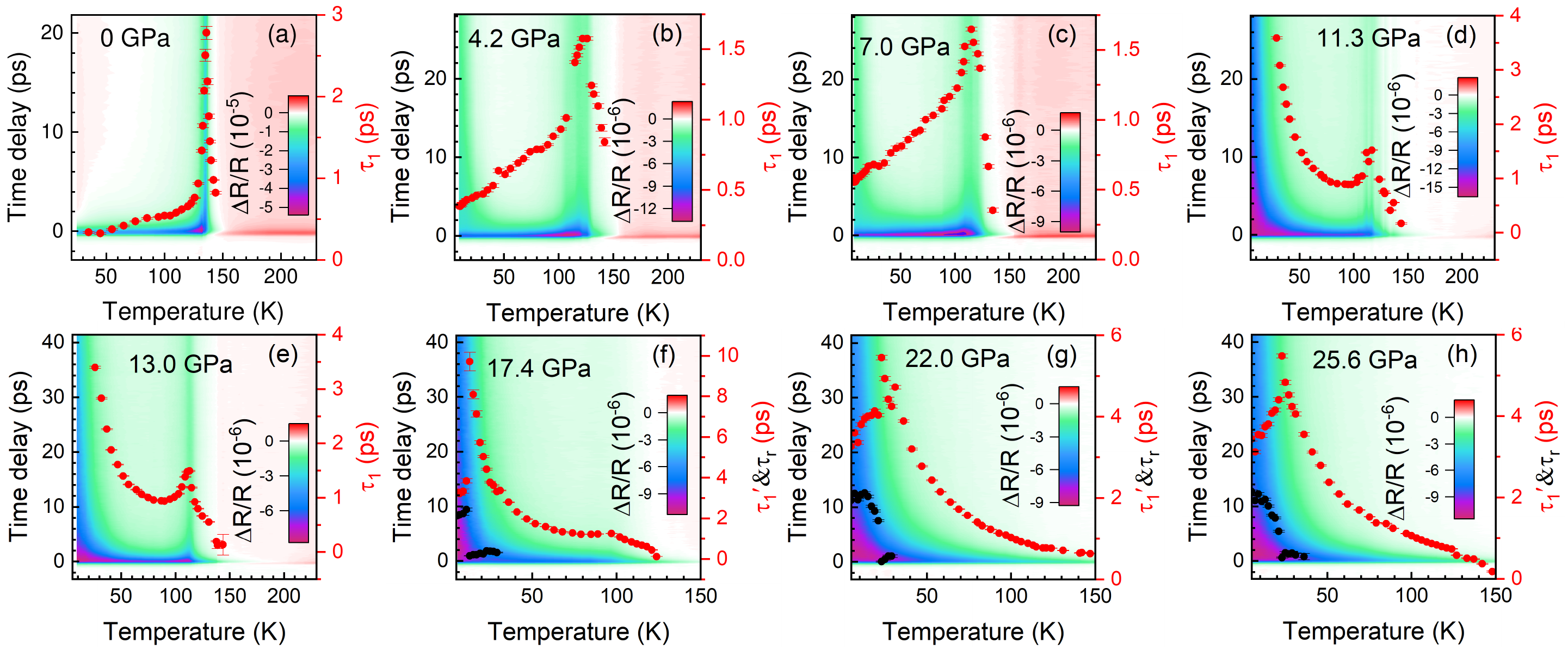}
	\caption{\textbf{The evolution of the density wave instability and emergence of superconductivity-like signal under various pressures.} The color plot of temperature-dependent transient reflectivity of La$_{4}$Ni$_{3}$O$_{10}$ under (a) 0 GPa; (b) 4.2 GPa; (c) 7.0 GPa; (d) 11.3 GPa; (e) 13.0 GPa; (f) 17.4 GPa; (g) 22.0 GPa; (h) 25.6 GPa. The red circle is the fitting value of lifetime $ \tau_{1} $ or $\tau_{1}^{\prime}$, while the black circle refers to the fitting value of rise time $ \tau_{r} $. Error bars refer to the standard error in the exponential fitting. The Phonon bottleneck effect are clearly observed from 0 GPa to 13 GPa with the peak of $ \tau_{1} $ decreasing from 136 K to 110 K. A slow rise component appears at 17.4 GPa and persists up to 25.6 GPa with the peak of $\tau_{1}^{\prime}$ increasing from 17 K to 23 K. The sign change of transient reflectivity from positive to negative occurs around 150 K and persists with increasing pressure without significant temperature variation.   
	}
	\label{Figure2}
\end{figure}

To clearly show the pressure evolution of DW and SC, the data from all temperature-dependent transient reflectivity $\Delta R/R$ at various pressures from 0 to 25.6 GPa are presented as a color map in Fig. \ref{Figure2}. As shown in Fig. \ref{Figure2}(a), a divergent behavior appears near $T_{DW}$ = 136 K for $P$ = 0 GPa. With increasing pressure, this divergent behavior in Fig. \ref{Figure2}(b)-(e) slowly moves to lower temperature at a rate of about -2 K/GPa (see the Supplementary Fig. 5) and then suddenly vanishes in Fig. \ref{Figure2}(f)-(h), indicating $T_{DW}$ slowly decreases to lower temperature and disappears above 17.4 GPa. Using the ratio of $\tau_{1}(T_{DW})$ to the $\tau_{1}(\mathrm{20 K})$ as a mark of the amplitude of divergence, we find that with increasing pressure, the amplitude of $\tau_{1}$ divergence gradually weakens. Therefore, we conclude that, although the phase transition temperature $T_{DW}$ reduces slowly with pressure, the amplitude of divergence in the relaxation time $\tau_{1}$ decreases rapidly as the pressure increases. Above 17.4 GPa, the DW gap feature disappears completely and then a SC-like signal appears. The transition temperature $T_{c}$ of SC, determined from the temperature at which $\tau_{1}^{\prime}$ diverges, are about 17 K, 22 K and 23 K at 17.4 GPa, 22.0 GPa and 25.6 GPa, respectively. It is worth to note that the dynamics is still highly temperature dependent far above $T_c$ at P $>$ 17.4 GPa, with slowing down of relaxation with decreasing temperature, indicating that the normal state is unusual. Similar behaviour has been observed in electron-doped cuprates \cite{PhysRevLett.110.217002}, which is a characteristic of short-range antiferromagnetic correlations. Further systematic fluence and temperature dependent measurement is necessary to define a certain characteristic temperature ($T^*$) below which the relaxation process is fluence dependent and to uncover an intimate connection between $T^*$ and the defined temperature by other probes. 

\begin{figure}
	\includegraphics[width=0.9\columnwidth]{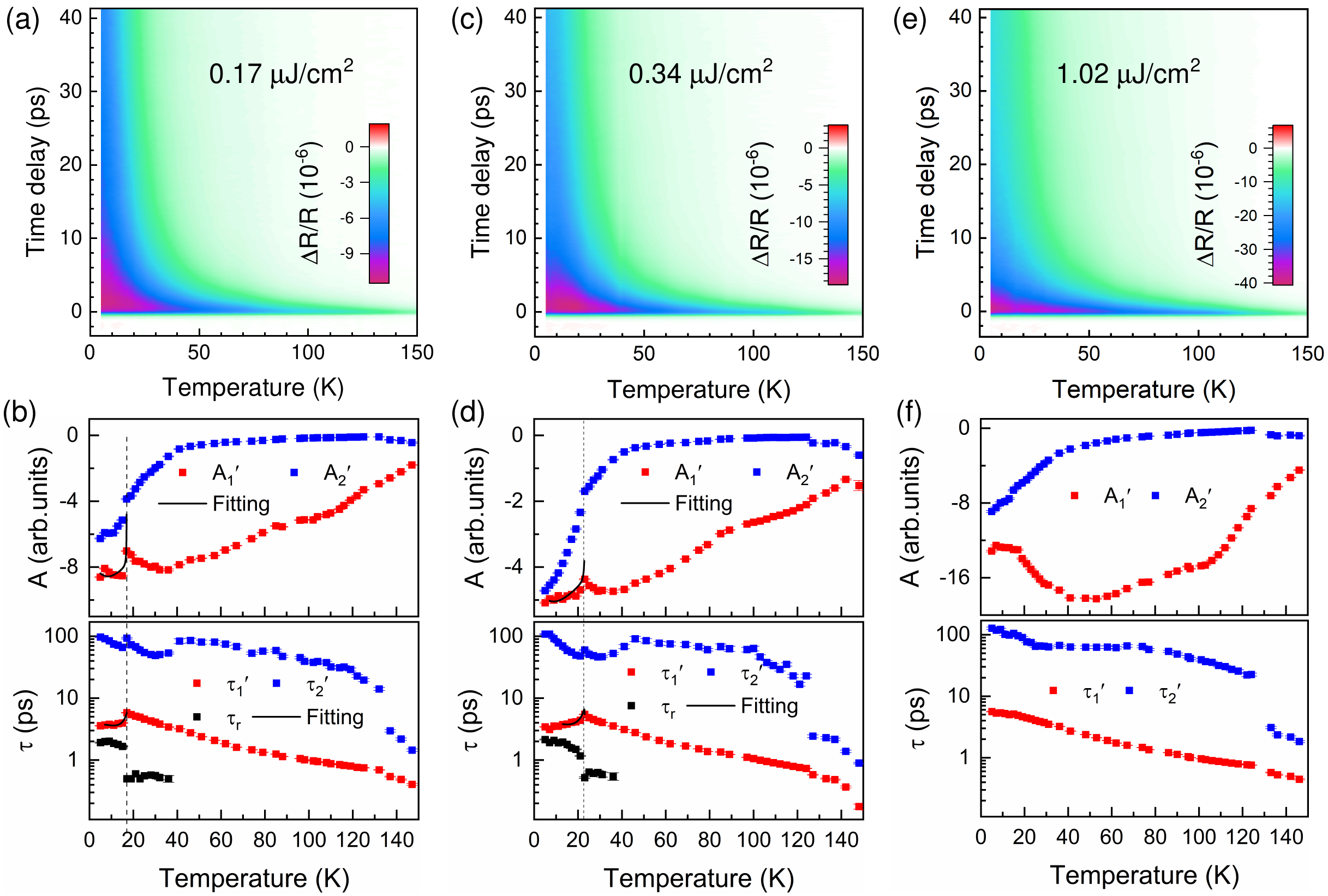}
	\caption{\textbf{The temperature and fluence dependence of the transient reflectivity of La$_{4}$Ni$_{3}$O$_{10}$ at 25.6 GPa.} (a) 0.17 $\mu$J/cm$^2$; (c) 0.34 $\mu$J/cm$^2$; (e) 1.02 $\mu$J/cm$^2$. The amplitude $A$ ($A_{1}^{\prime}$, $A_{2}^{\prime}$) and lifetime $\tau$ ($\tau_{1}^{\prime}$, $\tau_{2}^{\prime}$, $\tau_{r}$) were obtained and shown in (b), (d), (f) by fitting the curves in (a), (c), (e) with double-exponential function. Error bars refer to the standard error in the exponential fitting. The amplitude of transient reflectivity increases with increasing fluence, while the rise component decreases and disappears with fluence up to 1.02 $\mu$J/cm$^2$. The phonon bottleneck effect is observed in $A_{1}^{\prime}$ and $\tau_{1}^{\prime}$ at 0.17 $\mu$J/cm$^2$ and 0.34 $\mu$J/cm$^2$. The black solid lines in (b) and (d) represent the fitting results using Rothwarf-Taylor model. The black dashed line refers to the position of divergence temperature which is located at 23 K at 0.17 $\mu$J/cm$^{2}$ and 18 K at 0.34 $\mu$J/cm$^{2}$, respectively. 
	}
	\label{Figure3}
\end{figure}

As is well known, SC is usually sensitive to fluence and tends to degrade under high fluence \cite{PhysRevLett.95.117005,PhysRevLett.105.027005,PhysRevB.84.104518,PhysRevB.72.014544}. We measured temperature-dependent $\Delta R/R$ at different fluences under a fixed pressure of 25.6 GPa. Figure \ref{Figure3} presents the experimental results for fluences of 0.17 $\mu$J/cm$^{2}$ (a), 0.34 $\mu$J/cm$^{2}$ (c) and 1.02 $\mu$J/cm$^{2}$ (e). The results show that the amplitude of $\left| \Delta R/R \right|$ increases without saturation up to a fluence of 1.02 $\mu$J/cm$^{2}$. The area of purple region decreases significantly with fluence, implying a reduction in the photo-induced rise region. Furthermore, the experimental data in Fig. \ref{Figure3}(a) and \ref{Figure3}(c) are well-fitted using Eq. (2), whereas the data for higher fluence shown in Fig. \ref{Figure3}(e) fit well with Eq. (1). The corresponding fitting parameters are plotted in Fig. \ref{Figure3}(b), \ref{Figure3}(d) and \ref{Figure3}(f). Fig. \ref{Figure3}(b) shows that there is an obvious divergence in $\tau_{1}^{\prime}$ near 23 K, marked by black dashed line. With increasing fluence up to 0.34 $\mu$J/cm$^{2}$, this divergence survives, but the transition temperature decreases to around 18 K. Upon further increasing fluence to 1.02 $\mu$J/cm$^{2}$, this divergence becomes indistinguishable, indicating that SC signal disappears completely beyond this fluence. 

\begin{figure}
	\includegraphics[width=0.9\columnwidth]{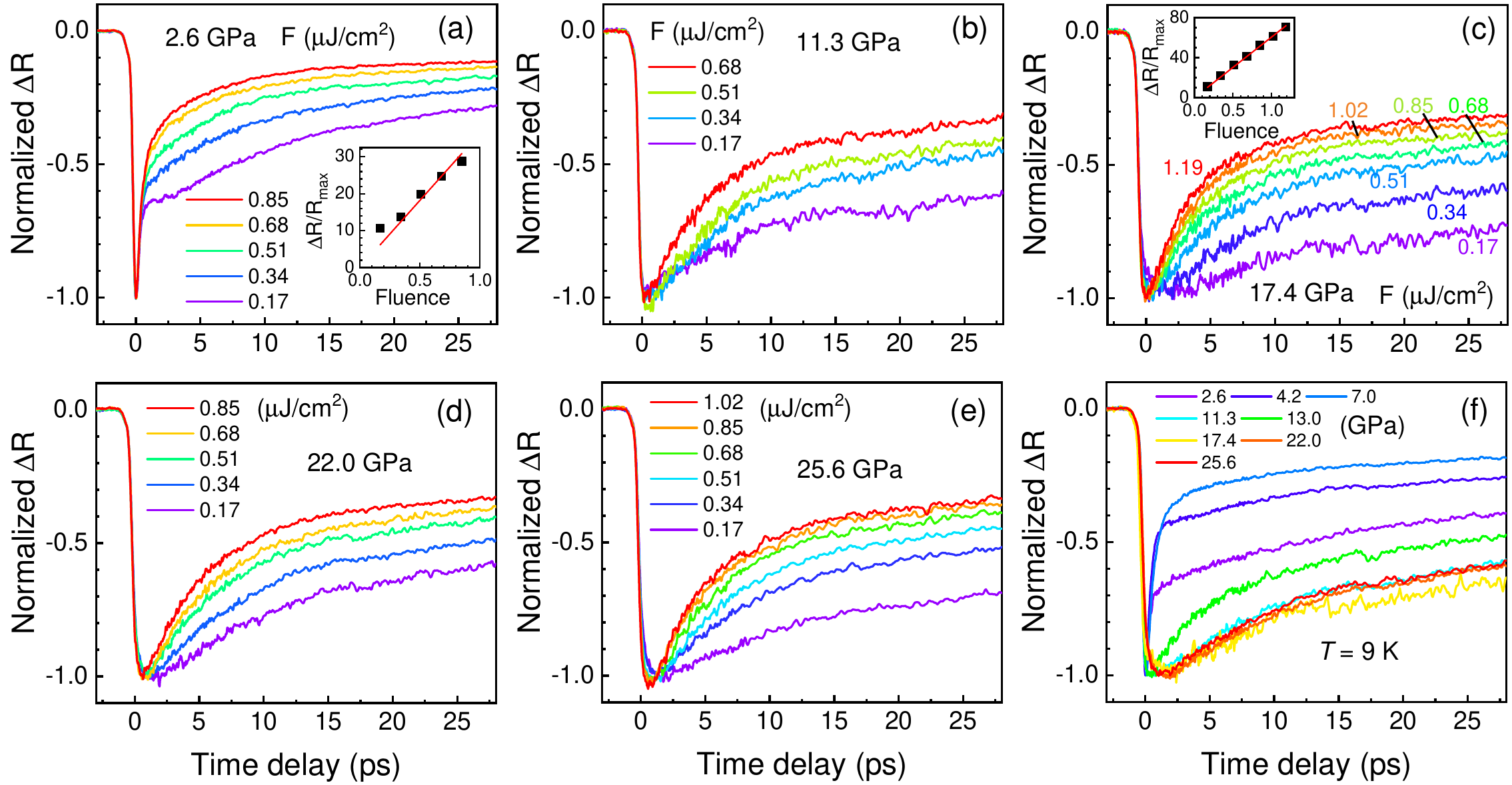}
	\caption{\textbf{Normalized $\Delta R$ versus fluence under various pressures at 9 K.} (a) 2.6 GPa; (b) 11.3 GPa; (c) 17.4 GPa; (d) 22.0 GPa; (e) 25.6 GPa. (f) Normalized $\Delta R$ versus pressure at the lowest fluence of 0.17 $\mu$J/cm$^2$ and 9 K. The inset of (a) and (c) shows the maximum of $\Delta R/R$ as a function of fluence, the solid red curve represents the linear fitting result. The $\Delta R/R$ increases linearly with fluence, but the relaxation process changes slightly with fluence up to 1.02 $\mu$J/cm$^2$, as shown in (c) and (e). A slow rise component appears at 17.4 GPa and persists to 25.6 GPa as shown in (f).}
	
	\label{Figure4}
\end{figure}

To further visualize the evolution of DW and SC under pressures, we plotted the normalized $\Delta R/R$ versus fluence under various pressures at a fixed temperature 9 K, as shown in Fig. \ref{Figure4}(a)-(e). It can be observed that all the relaxation processes become faster with increasing fluence. In addition, at low pressures (2.6 GPa and 11.3 GPa), the decay process lacks a slow rise component. However, at high pressures (17.4 GPa, 22.0 GPa and 26.0 GPa), a slow rise component is clearly observed and disappears rapidly as the fluence increases. These features are very similar to the decay processes observed in cuprate and iron-based superconductors below $T_c$ \cite{PhysRevB.95.115125,PhysRevLett.110.217002,PhysRevLett.95.117005,PhysRevLett.105.027005,PhysRevB.84.104518,PhysRevB.72.014544}. The inset of Figs. 4(a) and (c) shows that the maximum of $\left\vert\Delta R/R\right\vert$ increases linearly with fluence up to 1.02 $\mu$J/cm$^{2}$, indicating that the fluence of 0.17 $\mu$J/cm$^{2}$ we used is in the weak perturbation limit. Figure \ref{Figure4}(f) displays the normalized $\Delta R/R$ versus pressure at the lowest fluence of 0.17 $\mu$J/cm$^{2}$ and 9 K. We can clearly see that the relaxation process changes significantly with increasing pressure. A slow rise process appears at 17.4 GPa and persists up to the highest pressure of 25.6 GPa. The critical pressure for the appearance of superconductivity is close to that of the structural phase transition \cite{Zhu_2024,Li2024Stru}.

\begin{figure}
	\includegraphics[width=0.5\columnwidth]{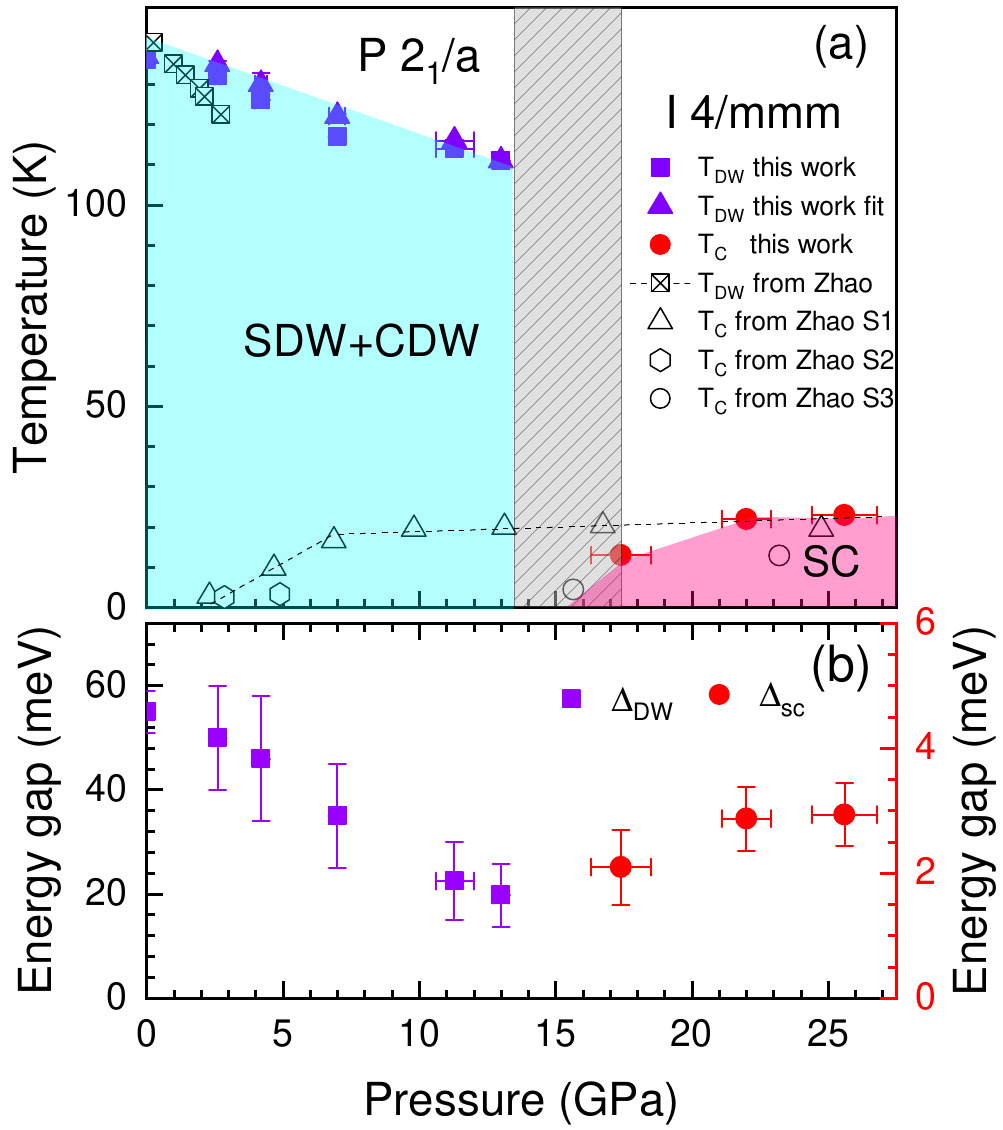}
	\caption{\textbf{Energy gap and temperature vs pressure phase diagram.} The purple square in (a) refers to the transition temperature of density wave $T_{DW}$ obtained from the phase diagram of color plot in Fig. 2. The purple triangle refers to $T_{DW}$ obtained from fit using RT model. The red circle in (a) represents the transition temperature of superconductivity $T_{c}$ obtained by RT model fitting. The hollow squares refer to $T_{DW}$ determined by high pressure transport measurements from Refs. \cite{Zhu_2024}. The hollow triangle, hexagon and circles represent to $T_{c}$ from Ref. \cite{Zhu_2024}. The black shadow refers to the range of critical pressure of structural transition from the monoclinic to the tetragonal phase. The density wave gap $\Delta_{DW}$ and possible superconducting gap $\Delta_{SC}$ in (b) are obtained by the fit with RT model. The vertical error bars refer to the standard error of the fit. The horizontal error bars indicate the mean pressure change from room temperature to 10 K.}
	\label{Figure5}
\end{figure}

Fig. \ref{Figure5}(a) summarizes the extracted $T_{DW}$ and $T_{c}$ in the temperature-pressure phase diagram. At low pressures, the DW transition $T_{DW}$ is gradually suppressed from 136 K at ambient pressure to 110 K at 13 GPa.  $T_{DW}$ suddenly becomes indistinguishable at 17.4 GPa. Then superconductivity appears after density wave collapse. The superconducting transition temperature $T_{c}$ appears near 17 K at 17.4 GPa and increases to 23 K at 25.6 GPa. This discontinuous transition sheds light on the role of repulsive interactions between DW and SC. According to previous x-ray diffraction data under high pressure, a structural transition from monoclinic to tetragonal phase occurs near 15 GPa \cite{Zhu_2024,Li2024Stru}. Therefore, the sudden disappearance of the density wave is closely related to the structural transition. In other words, the DW is stabilized exclusively by the monoclinic structure. In addition, the observation of a single DW gap under pressure indicates that the SDW and CDW components remain coupled as pressure increases, rather than decoupling which differs from that of the phase diagram of La$_{3}$Ni$_{2}$O$_{7}$\cite{2025unravelingspindensitywave,2025identicalsuppressionspincharge}.\\
According to previous reports, the DW transition is suppressed gradually under pressure \cite{Zhu_2024,Li__2024,zhang2024}. Refs. \cite{Zhu_2024} and \cite{Li__2024} show that the DW becomes indistinguishable above 10 GPa, whereas Ref. \cite{zhang2024} indicates that the DW still exists at 20 GPa. In Refs. \cite{Nagata_2024} and \cite{PhysRevB.109.144511}, the transport curve of La$_{4}$Ni$_{3}$O$_{10}$ does not show the characteristic features of a DW. However, all these studies reported signals of superconductivity under high pressure with different critical pressures. In our experiment, the critical pressure for the appearance of superconductivity is around 17.4 GPa and the maximum of $T_{c}$ we observed is around 23 K, lower than 30 K in Ref. \cite{Zhu_2024} and 36 K in Ref. \cite{Nagata_2024}. The lower $T_{c}$ is possibly caused by the lower applied pressure and the difference in oxygen content \cite{Nagata_2024,PhysRevB.109.144511}. In addition, the maximum $T_{c}$ of La$_{4}$Ni$_{3}$O$_{10}$ we measured is lower than that in La$_{3}$Ni$_{2}$O$_{7}$, which is consistent with other reports \cite{Sun2023,Zhu_2024}. According to our previous research, one possible explanation is the difference of electronic correlation strength \cite{xu2024origin}. Density functional theory shows that the stronger interlayer superconducting fluctuations in La$_{4}$Ni$_{3}$O$_{10}$ contribute to its lower $T_{c}$ \cite{QiongQin2024}. So far, there is no unified conclusion. The pump-probe spectroscopy of La$_{3}$Ni$_{2}$O$_{7}$ shows that the DW order exists throughout the entire pressure region, with a CDW phase occurring around 26 GPa \cite{Meng2024NC}. In addition, no SC signal was observed. However, in La$_{4}$Ni$_{3}$O$_{10}$, we clearly observed a SC-like signal. A possible explanation is that the fluence we used is small enough and the SC volume fraction in La$_{4}$Ni$_{3}$O$_{10}$ is large enough to detect the superconducting signal.\\
Having established the temperature-pressure phase diagram of La$_{4}$Ni$_{3}$O$_{10}$, now we attempt to extract the energy gaps of the DW and SC at the selected pressures, which usually is essential for understanding the mechanism of DW and SC. The Fig. \ref{Figure1}(d) and Supplementary Figs. 7 and 8 show that near the transition temperature, the amplitude of $A_{1}$ ($A_{1}^{\prime}$) sharply declines and $\tau_{1}$ ($\tau_{1}^{\prime}$) exhibits a diverge behavior. In the weak perturbation limit, where the density of photoexcited QPs is small compared to the density of thermally excited QPs, assuming the gap value follows the BCS-like behavior as $\Delta(T) = \Delta_{0}\sqrt{\left( 1 - T / T_{c} \right)}$, the temperature dependent $A$  and $\tau$ can be fitted with the following expressions \cite{PhysRevLett.19.27,PhysRevB.59.1497,PhysRevB.104.165110,PhysRevB.101.205112}:
\begin{equation}
	A(T) \propto \frac{\left( \Phi / \left( \left\lbrack \Delta(T) + \left( k_{B}T \right) / 2 \right\rbrack \right) \right)}{\left( 1 + \Gamma\sqrt{\left( 2k_{B}T/\pi\Delta(T) \right)}e^{ -\Delta(T) / \left( k_{B}T \right) } \right)}
\end{equation}  
and
\begin{equation}
	\tau(T) \propto \frac{ln\left\lbrack g + e^{ -\Delta(T)/\left( k_{B}T \right)}\right\rbrack}{\Delta(T)^{2}}
\end{equation}  
where $\Phi$ is the pump fluence, $ k_{B} $ is the Boltzmann constant,  $g$ and $\Gamma$ are the independent fitting parameters. The temperature-dependent $A_{1}$ ($A_{1}^{\prime}$) and $\tau_{1}$ ($\tau_{1}^{\prime}$) as well as their fitting curves under various pressures are plotted in the Supplementary Figs. 7 and 8. The extracted gap values for the density wave $\Delta_{DW}$ and superconductivity $\Delta_{SC}$ are summarized in Fig. \ref{Figure5}(b). The DW gap $\Delta_{DW}$ decreases from 55 meV at ambient pressure to 19 meV at 13 GPa. The effective DW gap energy is dramatically suppressed under applied pressure, whereas the suppression of $T_{DW}$ is relative modest. The disproportionate suppression of DW gap energy compared to $T_{DW}$ under pressure may reflect a redistribution of gap anisotropy or competing quantum fluctuations. Further momentum-resolved studies are needed to disentangle these effects. With increasing pressure, the DW gap closes, and SC appears. The gap of SC $\Delta_{SC}$ reaches 3 meV at 25.6 GPa and the value of a ratio $2\Delta_{SC}/k_{B}T_{c}$ is around 3.0. This ratio is close to the observed value in the weak coupling s-wave superconductors and some electron-doped cuprates \cite{dagan2007dirty,shan2008weak}, but is lower than the weakly coupled d-wave gap of 4.3. Previous studies have reported the detection of three superconducting gaps in La$_{3}$Ni$_{2}$O$_{7}$, including a small d-wave gap and two large s-wave gaps \cite{Liu2025}. However, our measurements of La$_{4}$Ni$_{3}$O$_{10}$ revealed only a single superconducting gap. Additionally, the gap-to-$T_{c}$ ratio of La$_{4}$Ni$_{3}$O$_{10}$ is similar to the minimum gap-to-$T_{c}$ ratio observed in La$_{3}$Ni$_{2}$O$_{7}$ \cite{Liu2025}. According to prior theoretical models, La$_{4}$Ni$_{3}$O$_{10}$is expected to exhibit a $s^{\pm}$ wave pairing symmetry similar to that of La$_{3}$Ni$_{2}$O$_{7}$ \cite{PhysRevLett.133.136001,PhysRevLett.131.236002}. Therefore, the superconducting gap we observed in La$_{4}$Ni$_{3}$O$_{10}$ may correspond to the minimum gap. The potential existence of larger superconducting gaps in La$_{4}$Ni$_{3}$O$_{10}$ warrants further investigation.  \\
Recently, dynamical mean-field theory has demonstrated that the correlated Fermi surfaces of high-pressure La$_{4}$Ni$_{3}$O$_{10}$ exhibit multiple in-plane nesting effects, which suggests competing spin and charge stripe orders \cite{PhysRevB.109.235123}. In addition, Refs. \cite{LI2025180} and \cite{lu2024super} show that the electron-phonon coupling strength is too weak to trigger superconductivity alone. The theoretical work also shows that, with the suppression of CDW and SDW orders under pressure, spin fluctuations increase sharply, highlighting their crucial role in the emergence of  superconductivity in La$_{4}$Ni$_{3}$O$_{10}$, similar to La$_{3}$Ni$_{2}$O$_{7}$ \cite{lu2024super,PhysRevB.108.L140505,PhysRevLett.131.236002,PhysRevB.109.235126}. The phase diagram of La$_{4}$Ni$_{3}$O$_{10}$ based on our ultrafast optical spectroscopic measurements (Fig. \ref{Figure5}) also shows that the suppression of DW is crucial to the emergence of superconductivity. Unlike cuprates, which exhibit a single $3d_{x^2-y^2}$ orbital near Fermi level, nickelates have two orbitals \cite{JPSJ.64.3179,PhysRevB.50.438,PhysRevLett.133.136001,PhysRevB.108.L140505,PhysRevLett.131.236002}. Nickelates also exhibit stronger interlayer spin coupling and weaker intralayer coupling compared to cuprates in the superconducting phase. The multi-orbital nature and strong electronic correlations in nickelates are critical for the emergence of superconductivity. A random-phase approximation based on a six-orbital tight-binding model shows that the gap function of La$_{4}$Ni$_{3}$O$_{10}$ exhibits an $s^{\pm}$ pattern, with the strongest pairing gap located at the $\mathit{\Gamma}$- and $M$-points with opposite signs  \cite{PhysRevB.110.L180501}. Indeed, the pairing symmetry of superconductivity in nickelates is sensitive to the Ni-$e_{g}$ crystal field splitting energy $\mathit{\Delta}$ \cite{Xia2025}. A decrease of $\mathit{\Delta}$ can change the pairing symmetry from $s^{\pm}$ wave to $d_{xy}$ wave. At low temperatures where the thermal excited QPs can be neglected, the measured amplitude of the $\Delta R/R$ is proportional to photoexcited QPs density ($ n $), while the absorbed laser fluence is proportional to the total energy stored in the QP system. According to previous experimental results \cite{PhysRevB.70.014504,PhysRevLett.105.027005} and theoretical study \cite{Fluence_theory}, for fully gapped excitations, $ n $ is proportional to the total energy absorbed, resulting in a simple relationship $\Delta R/R \propto \Phi$. In the presence of a line node, where the density of states depends linearly on energy, $ n $ should vary with the total stored energy in a sublinear manner $\Delta R/R \propto \Phi^{2/3}$. A clear linear dependence is observed at 17.4 GPa, indicating that the photoinduced QPs originate entirely from fully gapped excitations. Due to the multiband nature of nickelates, line node should also exist near Fermi surface. However, the exact pairing symmetry in nickelates needs further investigation with the energy and momentum-sensitive techniques.

In conclusion, we have constructed an in-situ low-temperature high-pressure ultrafast pump-probe spectroscopy setup to investigate the pressure evolution of density wave and superconductivity. The $T_{DW}$ decreases at a rate of -2 K/GPa with pressure and suddenly disappears above 17.4 GPa where structural transition appears. It means that the density wave only coexists with the monoclinic crystal structure and was destroyed after structural phase transition. Notably, a superconducting-like signal emerges after density waves collapse, indicating a competitive relationship between density waves and superconductivity. Moreover, we provide for the first time the experimental results on the energy gaps of density wave and superconducting phase of La$_{4}$Ni$_{3}$O$_{10}$ under pressure. The observation of a single DW gap under pressure indicates that the SDW and CDW components remain coupled as pressure increases, rather than decoupling. The superconducting gap we observed based on a symmetric \textit{T}-dependent gap assumption with a BCS-like \textit{T} dependence, falls within the weak-coupling limit. Our results provide crucial information for understanding the complex interplay among CDW, SDW and superconductivity in nickelates. In addition, this work advances the study of non-equilibrium quantum states under pressure using ultrafast pump-probe spectroscopy techniques.

\section{Methods}

\textbf{Crystal growth.} La$_{4}$Ni$_{3}$O$_{10}$ precursors for crystal growth were synthesized by solid-state reaction method. Stoichiometric amounts of La$_{2}$O$_{3}$  (Alfa Aesar, 99.99) and NiO (Alfa Aesar, 99.99) were mixed and thoroughly ground. Crystals of La$_{4}$Ni$_{3}$O$_{10}$ were successfully grown using a vertical optical-image floating zone furnace (100 bar model HKZ, SciDre). The precursor powder of La$_{4}$Ni$_{3}$O$_{10}$ was hydrostatically pressed into polycrystalline rods and sintered for 24 h at 1400 $^{\circ}$C to produce dense rods. High-quality La$_{4}$Ni$_{3}$O$_{10}$ single crystals were grown directly from the sintered rod at an oxygen pressure of 20 bar and a 5 kW xenon arc lamp. The structure characterization at ambient pressure and the resistivity measurements under high pressure can be found in Ref \cite{Li2024Stru}.

\textbf{Pressure effect.} A Be-Cu-type diamond anvil cell (DAC) with 500 microns flats was employed. A rhenium plate was used as a gasket to contain the sample and the pressure-transmitting medium. The rhenium plate was punched to create a hole with a diameter of 300 microns. Soft KBr powder was used as the pressure-transmitting medium due to its high transmittance and low absorption coefficient. In addition, soft KBr ensures sufficient contact between the sample and the diamond anvils, providing an excellent in-situ measurement environment. The La$_{4}$Ni$_{3}$O$_{10}$ crystal with a size of 200 $\mu$m$\times$120 $\mu$m was picked and loaded into the sample chamber. All the measurements were carried out in a $^{4}$He continuous flow cryostat (4.6 K $\leq$ $\mathit{T}$ $\leq$ 300 K).

\textbf{Pump probe experiments.} Time-resolved reflectivity experiments were conducted using an optical fiber oscillator with a center wavelength of 1560 nm, a repetition rate of 80 MHz, and a pulse duration of 120 fs. The pump and probe wavelengths are 780 nm and 1560 nm, respectively. Both beams are vertically polarized. The spot sizes of the pump and probe beams are focused to 120 $\mu$m and 80 $\mu$m on the sample, respectively. The vertically polarized pump beam was chopped at 733 kHz by an acousto-optic modulator to facilitate lock-in detection. The pump fluence was tuned to 0.2 $\mu$J/cm$^{2}$ while the probe fluence was reduced to only 10$\%$ of pump fluence. The sample signal is effectively distinguished from the signal of the diamond anvil surface using the principle of pinhole imaging with a lens. The reflectivity signal from the sample was detected using an amplified detector.

\textbf{In-situ pressure tuning and calibrations.} The ruby luminescence method was used for pressure measurement in the diamond anvil cell. The pressure $\mathit{P}$ is calculated by the formula $\mathit{P}=2.74(\lambda-\lambda_{0})$, where $\mathit{P}$ refers to pressure with a unit of GPa, $\lambda$ is the wavelength of ruby $\mathit{R_{1}}$ line, $\lambda_{0}$ is the reference wavelength of $\mathit{R_{1}}$ at ambient pressure \cite{Barnett1973,Syassen2008}. A pneumatic membrane was employed to achieve in-situ pressure tuning. The pressure in the pneumatic membrane, i.e., on the sample, could be continuously and precisely tuned by adjusting the pressure controller. The $^{4}$He gas was introduced into the pneumatic membrane to prevent freezing as the temperature decreases.

\medbreak
\textbf{Data availability} Source data are provided with this paper. Data generated or analyzed during this study are included in the Article and the Supplementary Information.

\section{References}

\bibliographystyle{apsrev4-2}
\bibliography{la4310}

\medbreak
\textbf{Acknowledgments} This work was supported by National Natural Science Foundation of China (Grant Nos. 12488201 by N.L.W., 12425404 by M.W.), National Key Research and Development Program of China (2024YFA1408700 by N.L.W, 2022YFA1403901 by N.L.W., 2021YFA1400201 by T.D.), the Guangdong Basic and Applied Basic Research Funds (Grant No. 2024B1515020040 by M.W.), Guangdong Provincial Key Laboratory of Magnetoelectric Physics and Devices (Grant No. 2022B1212010008 by M.W.), and Research Center for Magnetoelectric Physics of Guangdong Province (2024B0303390001 by M.W.). S. X. X. was also supported by Postdoctoral Science Foundation of China (Grant No. 2022M72071).

\medbreak
\textbf{Author contributions}: Shuxiang Xu, Tao Dong and Nanlin Wang conceived the project. The experimental setup and measurement program was constructed by Shuxiang Xu with the help of Hao Wang and Qiong Wu. Samples have been synthesized and characterized by Mengwu Huo and De yuan Hu under the supervise of Meng Wang. Dong Wu and Li Yue revised the manuscript carefully. Shuxiang Xu performed the high pressure pump-probe measurements. Shuxiang Xu, Tao Dong and Nanlin Wang analyzed the data and wrote the manuscript with inputs from all authors.
\medbreak
\textbf{Competing interests} The authors declare no competing financial interests. 
\medbreak
\textbf{Additional information} Supplementary information available. 
\clearpage

\clearpage

\end{document}